\documentclass[aps,prd,preprint,showpacs,showkeys]{revtex4-1}
\usepackage{mathrsfs}

\usepackage{bbm}
\usepackage{amsmath}

\usepackage{amssymb}
\usepackage{graphicx}

\begin{document}

\title{On Noether approach in the cosmological model with scalar and gauge fields: symmetries and the selection rule}

\author{Sun Zhang}
\affiliation{Key Laboratory of Dark Matter and Space Astronomy,
Purple Mountain Observatory, Chinese Academy of Sciences, Nanjing
210008, China
\\Joint Center for Particle, Nuclear Physics and
Cosmology (J-CPNPC), PMO-NJU, Nanjing 210008, China}

\begin{abstract}
In this paper, based on the works of Capozziello et al., we have
studied the Noether symmetry approach in the cosmological model with
scalar and gauge fields proposed recently by Soda et al. The
\emph{correct} Noether symmetries and related Lie algebra are given
according to the minisuperspace quantum cosmological model. The
Wheeler-De Witt (WDW) equation is presented after quantization and
the classical trajectories are then obtained in the semi-classical
limit. The oscillating features of the wave function in the cosmic
evolution recover the so-called Hartle criterion, and the selection
rule in minisuperspace quantum cosmology is strengthened. Then we
have realized now the proposition that Noether symmetries select
classical universes.
\end{abstract}


\maketitle

The Noether symmetry approach \cite{1} in general relativity and
cosmology is the endorsement of the crucial role of symmetries in
physics \cite{2,3,4,5,6}. As pointed out by Capozziello et al.
\cite{2,3,4}, such an approach can provide fundamental consideration
on the related models in terms of symmetries, and a selection rule
\cite{4} will come into being to recover the classical behaviors of
cosmic evolution due to the oscillating features of the cosmological
wave function. That is, the correlated region in the configuration
space of dynamical variables is selected and a classical observable
universe emerges. In the minisuperspace description, the so-called
Hartle criterion \cite{7} is applied to the solutions of the
Wheeler-De Witt (WDW) equation and selects the classical
trajectories \cite{2}.

Moreover, the vector field (or gauge field) has been taken into
account in cosmology, especially for the early universe
\cite{8,9,9a}. In the cosmological models with the vector field(s),
the anisotropic hair, perturbation, non-Gaussianity and other
aspects are studied extensively. Furthermore, if we are in the case
of gauge field, that is, the gauge symmetry is conserved in the
system, as proposed by Soda et al. \cite{8,9,9a}, the exact
anisotropic power-law solutions have been obtained when both the
potential function for the scalar field and the gauge kinetic
function are exponential type, and an attractor arises for a large
parameter region \cite{9}. The anisotropic hair related to the
so-called cosmic no-hair theorem and the mechanism of magnetogenesis
can also be discussed on this stage.

Furthermore, the Lie point symmetries of second-order partial
differential equations, including the Schr\"{o}dinger and the
Klein-Gordon equations have been studied in a general Riemannian
space by Paliathanasis et al. \cite{9.5a,9.5b}, and these symmetries
are related to the Noether point symmetries of the classical
Lagrangian. By employing the Lie symmetries of the WDW equation, a
general family of hyperbolic scalar field potentials is also
discussed within the framework of perfect fluid in Ref. \cite{9.5a},
and the exact solutions of the field equations could be obtained
based on the Lie symmetries of the WDW equation.

Then, if we want to discuss the Noether approach in cosmological
model with scalar and vector fields, we should start with the
following Lagrangian \cite{8,9}
\begin{equation} \label{z1}
S=\int d^4x\sqrt{-g}\left[\frac{M^2_p}{2}R-\frac{1}{2}(\partial_\mu
\phi)(\partial^\mu \phi)-V(\phi)-\frac{1}{4}f^2(\phi)F_{\mu
\nu}F^{\mu \nu}\right],
\end{equation}
where the flat Robertson-Walker background geometry is assumed, $g$,
$R$ and $a(t)$ are the determinant of the metric, the Ricci scalar
and the scale factor respectively, the reduced Planck mass takes
$M_p=1$, while $V(\phi)$ is a potential of the scalar field $\phi$
and $f(\phi)$ is the gauge kinetic function, a coupling between the
scalar field and the strength tensor $F_{\mu \nu}=\partial_\mu A_\nu
-\partial_\nu A_\mu$ for the vector field $A_\mu$. Such a model has
been studied recently in Noether symmetry approach by Vakili in Ref.
\cite{5}. Here, based on the works of Capozziello et al., we would
like to begin anew to provide the \emph{correct} Noether symmetries
and related Lie algebra thanks to the minisuperspace quantum
cosmological model. Then we try to obtain the solutions of the
Wheeler-De Witt (WDW) equation and present the classical
trajectories in the semi-classical limit. The Hartle criterion
should be applied and the selection rule is reinforced.

As in Refs. \cite{8,9}, we take the vector field $A_\mu$ along the
$x$-axis $A_\mu=(0,A_x(t),0,0)$, and the scalar field
$\phi=\phi(t)$, both of them are considered as homogeneous fields.
After some manipulations and do an integration by parts, a
point-like Lagrangian is obtained in the following form
\begin{equation} \label{z2}
\mathcal {L}= -3a \dot{a}^2+\frac{1}{2} a^3 \dot{\phi}^2+\frac{1}{2}
a f^2(\phi) \dot{A}^2-a^3 V(\phi).
\end{equation}
The configuration space for such a Lagrangian is $\mathcal
{Q}\equiv\left(a,\phi,A\right)$, and then cosmological dynamics can
be reached on such a three dimensional FRW minisuperspace.

According to Capozziello et al. \cite{2}, the vector field $X$
\begin{equation} \label{z3}
X=\alpha^i (q) \frac{\partial}{\partial q^i}+\dot{\alpha}^i (q)
\frac{\partial}{\partial \dot{q}^i}
\end{equation}
could be acted on a Lagrangian $\mathcal {L}$ defined on the tangent
space of configurations $T\mathcal {Q}$, then the Lie derivative of
$\mathcal {L}$ is
\begin{equation} \label{z4}
L_X \mathcal {L}=X \mathcal {L}=\alpha^i (q)\frac{\partial \mathcal
{L}}{\partial q^i}+\dot{\alpha}^i (q) \frac{\partial \mathcal
{L}}{\partial \dot{q}^i}.
\end{equation}
If $L_X \mathcal {L}=0$, we call $X$ a symmetry for the dynamics of
$\mathcal {L}$. It is clear that $\mathcal {L}$ in Eq. (\ref{z2})
does not depend on $A$, which is then cyclic. So we have the
symmetry definitely
\begin{equation} \label{z5}
X_3=\frac{\partial}{\partial A},
\end{equation}
that is, the related $\alpha^3$ is a constant and independent of
variables $a$, $\phi$ and $A$.

Considering the model Eq. (\ref{z2}) presented here, we have
\begin{equation} \label{z6}
X=\alpha^1  \frac{\partial}{\partial a}+\alpha^2
\frac{\partial}{\partial \phi}+\alpha^3  \frac{\partial}{\partial
A}+\dot{\alpha}^1 \frac{\partial }{\partial \dot{a}} +\dot{\alpha}^2
\frac{\partial }{\partial \dot{\phi}}+\dot{\alpha}^3 \frac{\partial
}{\partial \dot{A}}.
\end{equation}
If the Lie derivative of $\mathcal {L}$ vanishes along $X$, that is,
$L_X \mathcal {L}=0$ is satisfied, a constant of motion can be
deduced and the Noether theorem holds.

After a straightforward calculation, we get a quadratic expression
of $a$, $\phi$ and $A$. Then each coefficient should be zero and a
system of partial differential equations for $\alpha^i$ ($i=1,2,3$)
is obtained.
\begin{equation} \label{z7}
\alpha^1 +2a\frac{\partial \alpha^1}{\partial a}=0,
\end{equation}
\begin{equation} \label{z8}
3 \alpha^1+2a \frac{\partial \alpha^2}{\partial \phi}=0,
\end{equation}
\begin{equation} \label{z9}
-6\frac{\partial \alpha^1}{\partial \phi}+a^2 \frac{\partial
\alpha^2}{\partial a}=0,
\end{equation}
\begin{equation} \label{z10}
-6\frac{\partial \alpha^1}{\partial A}+f^2(\phi) \frac{\partial
\alpha^3}{\partial a}=0,
\end{equation}
\begin{equation} \label{z11}
a^2\frac{\partial \alpha^2}{\partial A}+f^2(\phi) \frac{\partial
\alpha^3}{\partial \phi}=0,
\end{equation}
\begin{equation} \label{z12}
\alpha^1 f(\phi)+2a\alpha^2 f'(\phi)+2a f(\phi) \frac{\partial
\alpha^3}{\partial A}=0,
\end{equation}
\begin{equation} \label{z13}
3\alpha^1 V(\phi)+a\alpha^2  V'(\phi)=0,
\end{equation}
where the prime is a derivative with respect to the scalar field
$\phi$.

To recover the symmetry Eq. (\ref{z5}), it requires that $\alpha^3$
is independent of variables $a$, $\phi$ and $A$. Then from Eq.
(\ref{z10}), $\partial \alpha^1/ \partial A=0$, and due to Eq.
(\ref{z7}),
\begin{equation} \label{z14}
\alpha^1 (a,\phi)=a^{-1/2}F(\phi),
\end{equation}
where $F$ is an arbitrary function dependent of $\phi$ only. From
Eq. (\ref{z11}), $\partial \alpha^2/ \partial A=0$, and due to Eq.
(\ref{z8}),
\begin{equation} \label{z15}
\alpha^2 (a,\phi)=-\frac{3}{2}a^{-3/2}\int F(\phi) d \phi.
\end{equation}
From Eq. (\ref{z12}), $ \alpha^1 f+2a \alpha^2 f'=0$, and
\begin{equation} \label{z16}
\frac{f'}{f}=-\frac{\alpha^1}{2a\alpha^2}.
\end{equation}
From Eq. (\ref{z13}), $ 3\alpha^1 V+a \alpha^2 V'=0$, and
\begin{equation} \label{z17}
\frac{V'}{V}=-\frac{3\alpha^1}{a\alpha^2}.
\end{equation}
Then from Eq. (\ref{z9}),
\begin{equation} \label{z18}
F'-\frac{3}{8}\int F d\phi=0,
\end{equation}
we can obtain the following solution for $F(\phi)$
\begin{equation} \label{z19}
F(\phi)=b_1 e^{\omega\phi}+b_2 e^{-\omega\phi},
\end{equation}
where $b_1$ and $b_2$ are constants of integration, and
$\omega^2=3/8$. The expressions for $\alpha^1$ and $\alpha^2$ are
also found
\begin{equation} \label{z20}
\alpha^1=a^{-1/2}  \left(b_1 e^{\omega\phi}+b_2
e^{-\omega\phi}\right),
\end{equation}
\begin{equation} \label{z21}
\alpha^2=-\frac{3}{2\omega} a^{-3/2} \left(b_1
e^{\omega\phi}-b_2e^{-\omega\phi}\right).
\end{equation}
It should be emphasized that there is no $A$ in the expressions of
$\alpha^1$ and $\alpha^2$, and the symmetry Eq. (\ref{z5}) is
recovered. From Eq. (\ref{z16})
\begin{equation} \label{z22}
\frac{f'}{f}=\frac{1}{8 \omega}\frac{b_1 e^{\omega\phi}+b_2
e^{-\omega\phi}}{b_1 e^{\omega\phi}-b_2 e^{-\omega\phi}},
\end{equation}
then the solution of $f(\phi)$ is
\begin{equation} \label{z23}
f(\phi)=\left(b_1 e^{\omega\phi}-b_2 e^{-\omega\phi}\right)^{1/3}.
\end{equation}
From Eq. (\ref{z17})
\begin{equation} \label{z24}
\frac{V'}{V}=\frac{3}{4 \omega}\frac{b_1 e^{\omega\phi}+b_2
e^{-\omega\phi}}{b_1 e^{\omega\phi}-b_2 e^{-\omega\phi}},
\end{equation}
then the solution of $V(\phi)$ is
\begin{equation} \label{z25}
V(\phi)=\left(b_1 e^{\omega\phi}-b_2 e^{-\omega\phi}\right)^2.
\end{equation}
So we reach a basis of symmetries on  $T\mathcal {Q}$,
\begin{equation} \label{z26}
X_1=a^{-1/2}e^{-\omega\phi}\frac{\partial}{\partial
a}+\frac{3}{2\omega}a^{-3/2}e^{-\omega\phi}\frac{\partial}{\partial
\phi}+\left(a^{-1/2}e^{-\omega\phi}\right)^.\frac{\partial}{\partial
\dot{a}}+\left(\frac{3}{2\omega}a^{-3/2}e^{-\omega\phi}\right)^.\frac{\partial}{\partial
\dot{\phi}},
\end{equation}
\begin{equation} \label{z27}
X_2=a^{-1/2}e^{\omega\phi}\frac{\partial}{\partial
a}-\frac{3}{2\omega}a^{-3/2}e^{\omega\phi}\frac{\partial}{\partial
\phi}+\left(a^{-1/2}e^{\omega\phi}\right)^.\frac{\partial}{\partial
\dot{a}}-\left(\frac{3}{2\omega}a^{-3/2}e^{\omega\phi}\right)^.\frac{\partial}{\partial
\dot{\phi}},
\end{equation}
\begin{equation} \label{z28}
X_3=\frac{\partial}{\partial A}.
\end{equation}
The corresponding constants of motion associated with the above
symmetries are
\begin{equation} \label{z29}
K_1=\frac{3}{2}a^{1/2}e^{-\omega\phi}\dot{a}-\omega
a^{-3/2}e^{-\omega\phi}\dot{\phi},
\end{equation}
\begin{equation} \label{z30}
K_2=\frac{3}{2}a^{1/2}e^{\omega\phi}\dot{a}+\omega
a^{3/2}e^{\omega\phi}\dot{\phi},
\end{equation}
\begin{equation} \label{z31}
K_3=af^2\dot{A}.
\end{equation}

It can be shown that all the symmetries are commuted with each
other, that is, the following Lie algebra is satisfied,
\begin{equation} \label{z32}
[X_i,X_j]=0,  \hspace{1.5cm} i,j=1,2,3.
\end{equation}
And the corresponding constants of motion also close the same
algebra in terms of Poisson bracket,
\begin{equation} \label{z33}
[K_i,K_j]=0,\hspace{1.5cm} i,j=1,2,3.
\end{equation}

According to Capozziello et al. \cite{2}, to identify the cyclic
variables, we make a transformation on the vector field Eq.
(\ref{z6}) (or Eq. (\ref{z26})) for $X_1$,
\begin{equation} \label{z34}
w=a^{3/2}e^{\omega\phi}, \hspace{1.5cm} z=a^{3/2}e^{-\omega\phi},
\hspace{1.5cm} A=A,
\end{equation}
to satisfied that
\begin{equation} \label{z35}
i_{X_1}dw=3, \hspace{1.5cm} i_{X_1}dz=0, \hspace{1.5cm} i_{X_1}dA=0.
\end{equation}
The configuration space is transformed from $\mathcal
{Q}\equiv\left(a,\phi,A\right)$ to $\mathcal
{\tilde{Q}}\equiv\left(w,z,A\right)$ correspondingly.

Then the Lagrangian Eq. (\ref{z2}) transformed into
\begin{equation} \label{z36}
\mathcal
{L}=-\frac{4}{3}\dot{w}\dot{z}-z^2+\frac{1}{2}z^{2/3}\dot{A}^2 ,
\end{equation}
and the conjugated momenta of dynamic variables $w$, $z$ and $A$ are
respectively
\begin{equation} \label{z37}
\pi_w=\frac{\partial \mathcal {L}}{\partial
\dot{w}}=-\frac{4}{3}\dot{z},
\end{equation}
\begin{equation} \label{z38}
\pi_z=\frac{\partial \mathcal {L}}{\partial
\dot{z}}=-\frac{4}{3}\dot{w},
\end{equation}
\begin{equation} \label{z39}
\pi_A=\frac{\partial \mathcal {L}}{\partial \dot{A}}=z^{2/3}\dot{A}.
\end{equation}

In the Hamiltonian dynamics, we should get the Hamiltonian firstly
\begin{equation} \label{z40}
\mathcal {H}=-\frac{3}{4}\pi_w\pi_z+\frac{1}{2}z^{-2/3}\pi_A^2 +z^2,
\end{equation}
and due to the Noether symmetries, the constants of motion are
\begin{equation} \label{z41}
\pi_w=\Sigma_0, \hspace{1.5cm} \pi_A=\Sigma_1.
\end{equation}
We can quantize the system as follows,
\begin{equation} \label{z42}
\pi_w\rightarrow \hat{\pi}_w=-i\partial_w, \hspace{1.3cm}
\pi_z\rightarrow \hat{\pi}_z=-i\partial_z, \hspace{1.3cm}
\pi_A\rightarrow \hat{\pi}_A=-i\partial_A,
\end{equation}
\begin{equation} \label{z42.5}
\mathcal {H}\rightarrow \mathcal {\hat{H}}(q^j,-i\partial_ {q^j}),
\end{equation}
then the Wheeler-De Witt (WDW) equation could be obtained
\begin{equation} \label{z43}
\left[-\frac{3}{4}(-i\partial_w)(-i\partial_z)+\frac{1}{2}z^{-2/3}(-i\partial_A)^2
+z^2\right]\mid\Psi\rangle=0.
\end{equation}
Separating the variables via the constants of motion,
\begin{equation} \label{z44}
\mid\Psi\rangle=\mid\Omega(w)\rangle \mid\chi(A)\rangle
\mid\xi(z)\rangle \propto e^{i\Sigma_0 w} e^{i\Sigma_1
A}\mid\xi(z)\rangle,
\end{equation}
the solution of $\xi(z)$ is found
\begin{equation} \label{z45}
\xi(z)=\exp\left[i\frac{2\Sigma^2_1}{\Sigma_0}z^{1/3}+i\frac{4}{9\Sigma_0}z^3\right].
\end{equation}
Such a wave function with the oscillating feature recovers the
so-called Hartle criterion \cite{7} and allows us to get a classical
observable universe.

If we now take the action (or the Hamiltonian principal function) as
\begin{equation} \label{z46}
S=\Sigma_0 \omega+\Sigma_1 A+
\frac{2\Sigma^2_1}{\Sigma_0}z^{1/3}+\frac{4}{9\Sigma_0}z^3,
\end{equation}
the corresponding momenta which conjugate the generalized
coordinates are obtained
\begin{equation} \label{z47}
\pi_w=\frac{\partial S}{\partial w}=\Sigma_0,
\end{equation}
\begin{equation} \label{z48}
\pi_z=\frac{\partial S}{\partial
z}=\frac{2\Sigma^2_1}{3\Sigma_0}z^{-2/3}+\frac{4}{3\Sigma_0}z^2,
\end{equation}
\begin{equation} \label{z49}
\pi_A=\frac{\partial S}{\partial A}=\Sigma_1.
\end{equation}
Separating the variables via above formulae, we can solve
Hamilton-Jacobi (HJ) equation in the semi-classical limit. The
classical trajectories in the configuration space, $\mathcal
{\tilde{Q}}\equiv\left(w,z,A\right)$, then can be expressed
\begin{equation} \label{z50}
z(t)=k_1t+k_2,
\end{equation}
\begin{equation} \label{z51}
A(t)=c(k_1t+k_2)^{1/3}+A_0,
\end{equation}
\begin{equation} \label{z52}
w(t)=c_1+c_2 (k_1t+k_2)^{1/3}+c_3 (k_1t+k_2)^3,
\end{equation}
where $k_2$, $A_0$ and $c_1$ are constants of integration, while
$k_1$, $c$, $c_2$ and $c_3$  are dependent of constants of motion
$\Sigma_0$ and $\Sigma_1$. If we go back to $\mathcal
{Q}\equiv\left(a,\phi,A\right)$, we can get the classical
cosmological solutions,
\begin{equation} \label{z54.5}
a(t)=\left[d_1(t-t_0)+d_2 (t-t_0)^{4/3}+d_3 (t-t_0)^4\right]^{1/3},
\end{equation}
\begin{equation} \label{z55}
\phi(t)=-\frac{1}{\omega} \ln \frac{k_1t+k_2}{\left[d_1(t-t_0)+d_2
(t-t_0)^{4/3}+d_3 (t-t_0)^4\right]^{1/2}},
\end{equation}
where $t_0$ and $d_i$ $(i=1,2,3)$ could be looked on as constants of
integration and should be determined by observations. Then the
oscillating regime is selected and the classical behaviors are
recovered. One of the main results is that the late time
acceleration could be obtained from the classical solution, and the
roll of scalar field may be more important than that of vector field
as discussed in Ref. [5]. As to the Theorem in Ref. [4], Noether
symmetries select classical universes. Such a picture has now been
realized here in the cosmological model with vector field(s)
proposed by Soda et al \cite{8,9}.

Moreover, the approach given here can be generalized naturally and
consistently to the case of phantom field \cite{10} and other cases
(such as the so-called quintom field) rather than other approaches.
If we take the following Lagrangian,
\begin{equation} \label{xx1}
S=\int
d^4x\sqrt{-g}\left[\frac{M^2_p}{2}R-\epsilon\frac{1}{2}(\partial_\mu
\phi)(\partial^\mu \phi)-V(\phi)-\frac{1}{4}f^2(\phi)F_{\mu
\nu}F^{\mu \nu}\right],
\end{equation}
then $\epsilon=+1$ is the standard case just discussed here, while
$\epsilon=-1$ is the phantom field case \cite{11}. Same approach
could be applied naturally though different behaviors are expected,
and related energy conditions are also to be discussed. Furthermore,
the related WDW equation is more complex, then we shall discuss the
Lie symmetries of the WDW equation as those in Ref. \cite{9.5a} to
identify the properties of the solution. All these should be
prepared systematically together with various other aspects and will
be placed somewhere else.

In summary, we have realized the proposition that Noether symmetries
select classical universes in the cosmological model with vector
(gauge) field(s). Based on the correct Noether symmetries, we give
the related Lie algebras for symmetries and constants of motion
according to the minisuperspace quantum cosmological model. We solve
the Wheeler-De Witt (WDW) equation after quantization and select a
subset of the solutions with oscillating behaviors. Then the
classical observable universe is recovered.

Finally, I would like to give a brief comment on arXiv:1410.3131
\cite{5}.

Though it is an interesting work, frankly, the symmetries in
arXiv:1410.3131 are incorrect.

Firstly and evidently, $\frac{\partial}{\partial A}$ is \emph{not} a
symmetry there, but in fact it is. While the constant of motion
related to this symmetry, called $P_{0A}$ or $P_{A}$, is used in the
calculation and plays a crucial role to obtain the final results and
conclusions. It is contradictive. We know that a constant of motion
should be corresponded to a symmetry. If there is no such a
symmetry, we cannot get the related constant of motion logically.

Secondly, other symmetries are also not correct, and no related
constants of motion can be given. The follow-up calculations seem
not to be directly based on these constants of motion according to
the method used in the paper, the incorrectness looks not so evident
like the one above. So the Lie Algebra is not given, in fact cannot
be given there, nor the WDW equation and so on.

Importantly, if the constant of motion used in the paper, called
$P_{0A}$ or $P_{A}$, could be obtained, it should be satisfied that,
based on its own derivation,
\begin{equation}
\dot{\gamma}=\frac{1}{2}\dot{a}\gamma_0 a^{-\frac{1}{2}}e ^{\mp
\omega\phi/3}\mp \frac{\omega \dot{\phi}}{3}\gamma_0
a^{\frac{1}{2}}e ^{\mp \omega\phi/3}=0,
\end{equation}
that is,
\begin{equation}
-\frac{3}{2} \frac{\dot{a}}{a} \pm \omega \dot{\phi}=0.
\end{equation}
While the one in the paper, eq. (38), is (the quantity $Q$ is set to
be zero)
\begin{equation}
-\frac{3}{2} \frac{\dot{a}}{a} \pm \omega \dot{\phi}+
3\frac{\dot{A}}{A} =0.
\end{equation}
It is clearly that $\frac{\dot{A}}{A}\neq 0$, so no any result like
$P_{0A}$ or $P_{A}$ given in the paper can be obtained.

The above equation is crucial to obtain the final results. Without
such an equation, no any of the final results shown in the paper
could be obtained. Taking some specific values for the constants,
the equations of the system become solvable and the results look
like correct ones. The results just happen to be superficially, the
calculations are incorrect.

The correct symmetries and the related Lie algebra are given in the
present paper, the cosmological variables are calculated by a
different method based on the works of Capozziello et al.

\begin{flushleft}
\textbf{Acknowledgements}
\end{flushleft}

The author is grateful to Professor Fan Wang, Dr. Wei Chen, Dr. Yu
Jiang, Dr. Feng-Yao Hou, Dr. Lei Feng, Dr. Yu-Peng Yang, Dr. Yan
Yan, Dr. Peng Dong for useful discussion and also to Mr. Yi-Qiao
Dong and Mr. K.B. Kim. This work is supported in part by the
National Natural Science Foundation of China (Grant Nos. 11373068,
10973039 and 10447114).

\end{document}